\documentclass[showpacs,preprintnumbers,amsmath,amssymb]{revtex4} 
\usepackage{graphicx}
\usepackage{dcolumn}
\usepackage{bm}
\usepackage{natbib}
\usepackage{amssymb,amsmath}
\usepackage{longtable}
\usepackage{txfonts}
\usepackage[colorlinks]{hyperref}
\bibpunct{(}{)}{;}{a}{}{,}

\makeatletter
\newcommand*{\rom}[1]{\expandafter\@slowromancap\romannumeral #1@}
\makeatother

\newcommand{\aap}{Astronom. and Astrophys.}
\newcommand{\pasa}{Publications of the Astronomical Society of Australia}

\newcommand{\aaps}{Astronom. and Astrophys. Suppl. Ser.}

\newcommand{\aj}{Astronom. J.}
\newcommand{\apjs}{Astrophys. J. Suppl.}

\newcommand{\mnras}{Monthly Notices Roy. Astronom. Soc.}

\newcommand{\pasp}{Publ. Astronom. Soc. Pacific}

\def\fm{\hbox{$.\!\!^m$}}

\def\degr{\hbox{$^\circ$}}

\begin{document}
\title{Physical and Dynamical Parameters of the Triple Stellar System: HIP 109951}
\author{\firstname{S.\,G.}~\surname{Masda}$^{1,2,}$}
\email{suhail.masda@gmail.com}
\affiliation{$^1$ Department of Physics, Dr. Babasaheb Ambedkar Marathwada University, Aurangabad-431001, Maharashtra, India.\\$^2$Department of Physics, Hadhramout University, PO Box:50511-50512, Mukalla, Yemen.}

\author{\firstname{J.\, A.}~\surname{Docobo}}
\email{joseangel.docobo@usc.es}
\affiliation{Observatorio Astron\'{o}mico Ram\'{o}n Mar\'{i}a Aller, Universidade de Santiago de Compostela, Avenida das Ciencias s/n, 15782 Santiago de Compostela, Spain.}
\affiliation{Real Academia de Ciencias de Zaragoza. Facultad de Ciencias.\\
	C/ Pedro Cerbuna 12. 50009 Zaragoza. Spain}

\author{\firstname{A.\,M.}~\surname{Hussein}}
\affiliation{Department of Physics, Al al-Bayt University, PO Box: 130040, Mafraq, 25113 Jordan.}

\author{\firstname{M.\,K.}~\surname{Mardini}}
\affiliation{Key Lab of Optical Astronomy, National Astronomical Observatories, Chinese Academy of Sciences, Beijing, China}
\affiliation{School of Astronomy and Space Science, University of Chinese Academy of Sciences,  Beijing, China}

\author{\firstname{H.\, A.}~\surname{ Al-Ameryeen}}
\affiliation{Observatorio Astron\'{o}mico Ram\'{o}n Mar\'{i}a Aller, Universidade de Santiago de Compostela, Avenida das Ciencias s/n, 15782 Santiago de Compostela, Spain.}
\author{\firstname{P.\,  P.}~\surname{ Campo}}
\affiliation{Observatorio Astron\'{o}mico Ram\'{o}n Mar\'{i}a Aller, Universidade de Santiago de Compostela, Avenida das Ciencias s/n, 15782 Santiago de Compostela, Spain.}
\author{\firstname{A.\,  R.}~\surname{ Khan}}
\affiliation{Department of Physics, Maulana Azad College, Aurangabad-431001, Maharashtra, India.}
\author{\firstname{J.\,  M.}~\surname{ Pathan}}
\affiliation{Department of Physics, Maulana Azad College, Aurangabad-431001, Maharashtra, India.}

\received{February 17, 2019}%
\revised{September 15, 2019}%


\begin{abstract}
The precise determination of the physical and dynamical parameters of the HIP 109951 triple star system (WDS J22161-0705AB) which is formed by the  A, Ba, and Bb components are presented. The binary nature of  component B was recently confirmed by studying the radial velocities. 
The analysis of the system follows  Al-Wardat's complex method for analyzing CVBS which employs Kurucz (Atlas9) line blanketed plane-parallel atmospheres simultaneously with an analytic method for dynamical analysis  (we used Docobo's method) to calculate the parameters of this triple system.
The result of our study yielded the following parameters: $T^{A}_{\rm eff}=5836\pm 80$ K, $R^{A}=1.090\pm0.039R_\odot$, $\log$g$^{A}=4.45\pm0.06$, $\mathcal{M}^{A}=1.05\pm0.16\mathcal{M}_\odot$;  $T^{Ba}_{\rm eff}=5115\pm 80$ K, $R^{Ba}=0.596\pm0.05R_\odot$, $\log$ g$^{Ba}=4.60\pm0.07$, $\mathcal{M}^{Ba}=0.83\pm0.16\mathcal{M}_\odot$,  $T^{Bb}_{\rm eff}=4500\pm 80$ K, $R^{Bb}=0.490\pm0.06R_\odot$, $\log$ g$^{Bb}=4.65\pm0.07$, and  $\mathcal{M}^{Bb}=0.67\pm0.16\mathcal{M}_\odot$ based on the revised \textit{Hipparcos} parallax. The orbital solution gave a total mass as  
$\mathcal{M}=2.59\mathcal{M}_\odot$ based on Gaia parallax and $\mathcal{M}=2.15\mathcal{M}_\odot$ based on the revised \textit{Hipparcos} parallax. The synthetic spectral energy distributions (SED) and synthetic stellar photometry of the entire system and individual components are given and compared with the available observational ones. Finally, the positions of the system components on the HR diagram and the evolutionary tracks are given and their formation and the evolution of the system are discussed.

\end{abstract}

\maketitle

\section{Introduction}

The study and analysis of physical and dynamical parameters of binaries is crucial for testing models of stellar formation and evolution. One of those parameters is the stellar mass which is of utmost importance in Astrophysics.
The stellar masses are fundamentally derived either from  a precise orbit calculation in the case of binary and triple systems or from the position of a specific star on the HR diagram. 
On the other hand, the estimation of the theoretical values of  masses and ages of binary systems, specifically solar-type stars on the main sequence with  accuracy comparable to direct methods, remains an arduous task \citep{2010MNRAS.401..695M,2015ApJ...812...96G}. One of the indirect methods makes use of positions of the components of the system on the Hertzsprung-Russell (HR) diagram.  Said method is explicitly dependent on the evolutionary tracks of the system  \citep{2000AAS..141..371G,2000yCat..41410371G}.

This paper presents a comprehensive analysis of the physical and dynamical parameters of the  HIP\,109951 system. This system is formed by three components: A, Ba, and Bb. First, the astrometrical satellite Hipparcos discovered the visual pair, AB, and Tokovinin \citep{2018AJ....156...48T}, studying the radial velocities of the system confirmed that the B component is binary. According to this research, the components, Ba and Bb have different magnitudes and the couple can be considered a single-lined spectroscopic binary, SB1. For this reason, it is probable that the difference of magnitude between their components is equal to or greater than 1.5. Its parallaxes were recently published by the Gaia project as $\varpi_{Gaia}=15.1176\pm 0.5342$ mas \citep{2018yCat.1345....0G} and the revised \textit{Hipparcos} parallax as $\varpi_{Hip}=16.09\pm 1.07$ mas \cite{2007AA...474..653V} which implicate a kinematic distance of $66.148\pm0.002$ and $62.15\pm0.004$ pc, respectively. Speckle interferometry revealed a separation between the main component, A, and the pair Ba-Bb close to $0.4$ arcsec and it yielded 17 relative position measurements after the first astrometric measurement by Hipparcos on 1991.25 (\textit{Hipparcos} Catalog ~\citep{1997yCat.1239....0E}).

The system is studied following Al-Wardat's complex method for analyzing close visual binary systems (CVBSs) \citep{2002BSAO...53...51A, 2007AN....328...63A, 2012PASA...29..523A} which uses the magnitude difference between the two components to build a synthetic spectral energy distribution (SED) for each component and then to combine them together in order to be comparable to the entire observational SED and photometry  in an iterated method. This approach was combined with Docobo's dynamical analysis \citep{1985CeMec..36..143D, 2012ocpd.conf..119D, 2018MNRAS.476.2792D} in order to determine the complete set of physical and dynamical parameters of the system. In addition, Al-Wardat's complex method was successfully applied to estimate the stellar parameters of several solar-type stars and sub-giant binary stars such as HIP\,70973, HIP\,72479, HIP\,11352, HIP\,11253, HD\,25811, HD\,375, Gliese\,150.2, Gliese\,762.1, HD\,6009, FIN\,350, COU1511, HIP\,105947, HIP\,14075 and HIP\,14230 ~\citep{2012PASA...29..523A, 2009AN....330..385A, 2009AstBu..64..365A, 2014AstBu..69..198A,  2014AstBu..69...58A, 2014AstBu..69..454A,  2014PASA...31....5A,2016RAA....16..166A,2017AstBu..72...24A,2018arXiv180203804M,2018JApA...39...58M}.

The first orbit of the system A-Ba,Bb was published by~\cite{2010AJ....139..205H} using the code of~\cite{2004AAS...204.0719M} with a period of 134 yr, a semi-major axis of 0.567\,arcsec, and a mass-sum of $2.98\pm0.91\,\mathcal{M}_\odot$ using the old \textit{Hipparcos} parallax of $15.04\pm1.52$\,mas~\citep{1997yCat.1239....0E}. The orbit was then modified by~\cite{2014AJ....147...62C} and \cite{2015ApJ...799....4R} using a completely different method. Their estimated mass-sums were $2.18\pm0.44\,\mathcal{M}_\odot$ and $3.35\pm6.52\,\mathcal{M}_\odot$, with periods of 80.57 and 44 years, respectively. The last orbit of the system was calculated by~\cite{2018AJ....156...48T} using high-resolution radial velocities and speckle measurements. Tokovinin \citep{2018AJ....156...48T} gave a mass sum of $2.58\,\mathcal{M}_\odot$ using a visual orbit with a period of 50.49 years and the Gaia parallax \cite{2018yCat.1345....0G}.

\section{Observational data}
The entire observational SED for the HIP\,109951 system was taken from~\cite{2003BSAO...55...18A} (Figure~\ref{aa2}). This SED  reveals significant information about the properties of the system and will be used as a reference for comparison with the synthetic ones. The observational SED was  obtained  using  a low-resolution grating ($325/4^{\circ}$ grooves/mm, $5.97$ {\AA}/px reciprocal dispersion) within the UAGS spectrograph on the 1\,m (Zeiss-1000) telescope at SAO-RAS-Russia and it covers the wavelength range from 3700\,\AA to 8000\,\AA.


Table~\ref{tab1} contains the fundamental data for HIP\,109951 from the SIMBAD database and Table~\ref{tab2} contains photometric data of the system from the NASA/IPAC, the \textit{Hipparcos} and the Tycho Catalogs \citep{1997yCat.1239....0E}, and Str\"{o}mgren \citep{1998AAS..129..431H}. Table~\ref{tab33} shows the magnitude differences ($\rm \Delta m$) between the two main components of the visual system along with filters used in the observation expressed in nm as well as a reference for each measurement from the Fourth Catalog of Interferometric Measurements of Binary Stars (INT4, \cite{2001AJ....122.3480H})~\footnote{http://www.usno.navy.mil/USNO/astrometry/optical-IR-prod/wds/int4}.

\begin{table}[ht]
	\caption{Fundamental data from SIMBAD for the HIP\,109951 system} \label{tab1}
	\begin{tabular}{@{}c|c|c}\hline\hline
		Property&  HIP\,109951& Source of data 		\\
		&  HD\,211276 &  \\
		\hline
		$\alpha_{2000}$ $^1$&   $22^{\rm h} 16^{\rm m} 06^{\rm s}565$& SIMBAD$^2$\\
		$\delta_{2000}$ $^3$ 	& $-07\degr05' 26.''62$& -\\
		SAO  &  145984& -\\
		Sp. Typ.   & G5&-
		\\
	E(B-V)      & $0.061\pm0.002$& NASA/IPAC$^{4}$
	\\	
	E(B-V)      & $0.052\pm0.002$& NASA/IPAC$^{5}$\\	
	$A_V$       &$0.19\pm 0.002$& NASA/IPAC$^{4}$
	\\
	$A_V$       &$0.16\pm 0.002$& NASA/IPAC$^{5}$
		\\
		\hline\hline
	\end{tabular}
	\\
	Notes.
	$^1$ Right Ascention,  	$^2$ http://simbad.u-strasbg.fr/simbad/sim-fid.\\
	$^3$ Declination,
	$^{4}$~\cite{1998ApJ...500..525S} and $^{5}$ \cite{2011ApJ...737..103S}.
\end{table}

\begin{table}[ht]
	\caption{Photometric data of the system, HIP\,109951, from the Hipparcos and the TYCHO
		catalogs, including TYCHO- $B_T$ and $V_T$ magnitudes, Str\"{o}mgren photometry, and trigonometric Hipparcos and Gaia parallaxes} \label{tab2}	
	\begin{tabular}{@{}c|c|c}\hline\hline
		Property&  HIP\,109951& Source of data 		\\
		&  HD\,211276 &  \\
		\hline
		$V_J(Hip)$  & $8.72$ & \cite{1997yCat.1239....0E}		\\
		$B_J$       & $9.43\pm0.03$& \cite{2000AA...355L..27H}		\\
		$(V-I)_J$&  $0.76\pm0.00$ & \cite{1997yCat.1239....0E}		\\				
		$(B-V)_J$&  $0.71\pm0.002$&\cite{1997yCat.1239....0E}		\\
		$B_T$  &  $9.58\pm0.03$ & \cite{2000AA...355L..27H}		\\			
		$V_T$ &  $8.82\pm0.02$ & \cite{2000AA...355L..27H}		\\
		$(u-v)_S$& $0.96\pm0.003$& \cite{1998AAS..129..431H} \\
		$(v-b)_S$& $0.68\pm0.003$& \cite{1998AAS..129..431H}\\
		$(b-y)_S$& $0.44\pm0.003$& \cite{1998AAS..129..431H}		\\
		$\varpi_{\rm Hip}$ (mas)      & $15.04\pm1.52$ & \cite{1997yCat.1239....0E}		\\
		$\varpi_{\rm Hip}$ (mas)    &  $16.09\pm1.07$ & \cite{2007AA...474..653V}		\\
		$\varpi_{\rm Gaia}$ (mas)    &  $15.12\pm0.534$ &  \cite{2018yCat.1345....0G} \\
		\hline\hline
	\end{tabular}
	\\
	
\end{table}

\begin{table}[ht]
	\begin{center}
		\caption{Magnitude difference between the components of the HIP 109951 system and available errors, along with filters used to obtain the observations}
		\label{tab33}
		\begin{tabular}{@{}c|c|c|c|c}
			\hline\hline
	Comp.		& $\triangle\rm m $& {$\sigma_{\Delta\rm m}$}& Filter ($\lambda/\Delta\lambda$) & Ref.$^\dagger$  \\
	vector	&	(mag) & &  ($~\rm nm$) &   \\
			\hline
	A--B	&	$1.83$ &   0.87  & $V_{Hip}:550/40 $&  \cite{1997yCat.1239....0E}  \\		
		&	$1.92$ &   0.04  & $545/30 $&  \cite{2005AA...431..587P}   \\						
		&	$1.87$ &  0.04   & $545/30 $&  \cite{2004AA...422..627B}   \\		
		&	$1.70$ &   0.13  &$648/41 $&  \cite{2004AJ....127.1727H}    \\
		&	$1.88$ &  0.08  & $545/30$ &   \cite{2006BSAO...59...20B}  \\		
		&	$1.53$ &  *  &$754/44 $ &  \cite{2008AJ....136..312H}  \\
		&	$1.88$ &  0.03  &$545/30 $&   \cite{2007AstBu..62..339B}  \\			
		&	$1.81$ & *  & $698/39 $  &  \cite{2008AJ....136..312H}  \\
		&	$1.70$ & *   & $698/39 $  &  \cite{2008AJ....136..312H}  \\
		&	$1.44 $ &  *    & $754/44 $  &   \cite{2010AJ....139..205H}     \\
		&	$1.94$ &   *   & $550/40 $  & \cite{2010AJ....139..205H}  \\
		&	$1.86$ &  0.10   &  $550/40 $  &  \cite{2012AJ....143...10H}        \\
		&	$1.76$ &   0.10  & $550/40 $  &  \cite{2012AJ....143...10H}       \\
			
		&	$2.00$ &   *    & $551/22 $  &  \cite{2010AJ....139..743T}      \\
		\hline
  Ba--Bb	&	$\geq1.50$ &      &   &       \\
			\hline\hline
		\end{tabular}
		\\
		$^*$ The errors are not given in INT4.\\
		
	\end{center}
\end{table}

\section{Analysis}
\subsection{Orbit}
In order to obtain the orbital solution, we have used Docobo's analytic method \citep{1985CeMec..36..143D, 2012ocpd.conf..119D, 2018MNRAS.476.2792D}. It is well known that this method is based on the selection of three base points and, with them, the family of relative orbits with corresponding apparent orbits that pass through the base points is generated. Solution selection can be based on different criteria although the most commonly used procedure is to calculate the root mean squares (RMS) of the residuals in the position angles and the separations. Of course, the weights of the observations are taken into account.

\begin{figure}[ht]
	\includegraphics[angle=0,width=14cm]{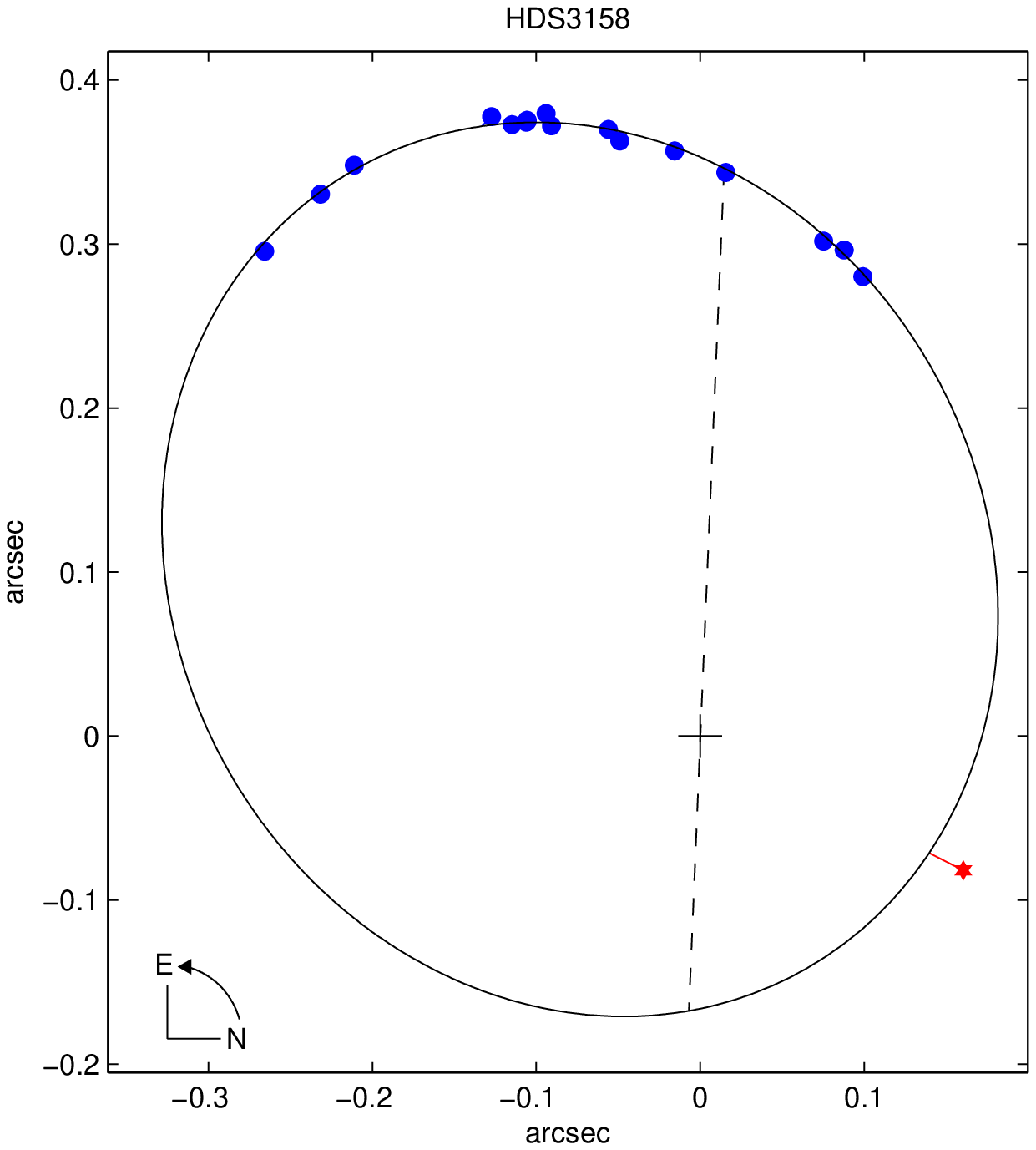}
	\caption{The apparent orbit of the  Ba,Bb subsystem around the primary, A. The relative orbit was calculated with Docobo's method using measurements from the INT4~\citep{2001AJ....122.3480H}. As ususal, the origin point represents the position of the primary component and the dashes indicate the line of the nodes. The red star represents the first \textit{Hipparcos} measurement.}\label{2}
\end{figure}

Tables~\ref{po2} and \ref{tabb} contain information about the residuals in $\theta$ and $\rho$ obtained with the new calculated orbit, the best orbital elements, and the RMS, respectively. The apparent orbit of the visual binary, A-Ba,Bb, is shown in Figure~\ref{2}.

\begin{table}[ht]
	\begin{center}
		\caption{Columns 1, 2, and 3 show the epochs and the relative position measurements taken from the INT4. The residuals, $\Delta \theta$ and $\Delta \rho$, obtained with the orbit presented in this paper, are listed in Columns 4 and 5. Column 6 contains the reference numbers. At the bottom, the rms in $\theta$ and $\rho$ are included}
		\small
		\label{po2}
		\centering
		\begin{tabular}{l|r|c|c|c|r}
			\hline\hline
			Epoch	& $\theta$ & $\rho$ & $\Delta \theta $ & $\Delta \rho $ & Ref.\\
			&(\degr) & ($''$) & (\degr) & ($''$) & \\
			\hline			
			1991.25& $333.0$ & $0.180$ & 0.0 &  0.024 & \cite{1997yCat.1239....0E} \\			
			1999.8152 & $70.50$ & $0.297$ &  0.8  &  0.002 & \cite{2004AA...422..627B} \\			
			2000.7672 & $73.5$  & $0.309$ & -1.4  &  -0.001 & \cite{2002AJ....123.3442H} \\
			2000.8726 & $76.0$  & $0.311$ &  0.6  &  0.000 & \cite{2006BSAO...59...20B} \\
			2003.6368 & $87.4$  & $0.344$ & -0.6  & -0.002 & \cite{2008AJ....136..312H} \\
			2004.8152 & $92.5$  & $0.357$ & -0.1  & -0.002 & \cite{2007AstBu..62..339B} \\
			2006.5173 & $98.6$  & $0.374$ & -0.3  &  0.000 & \cite{2008AJ....136..312H} \\
			2006.5174 & $97.7$  & $0.366$ & -1.2  & -0.008 & \cite{2008AJ....136..312H} \\			
			2007.8171 & $103.7$ & $0.383$ &  0.3  &  0.000 & \cite{2007AstBu..62..339B} \\
			2007.8199 & $103.9$ & $0.391$ &  0.5  &  0.008 & \cite{2007AstBu..62..339B} \\			
			2008.4723 & $105.7$ & $0.390$ &  0.2  &  0.003 & \cite{2012AJ....143...10H} \\
			2008.7018 & $107.1$ & $0.390$ &  0.8  &  0.002 & \cite{2012AJ....143...10H} \\
			2008.7670 & $105.8$ & $0.389$ & -0.7  &  0.000 & \cite{2010AJ....139..743T} \\
			2008.7670  & $105.7$ & $0.390$ & -0.8  &  0.001 & \cite{2010AJ....139..743T} \\			
			2009.7553 & $108.6$ & $0.399$ & -1.1  &  0.005 & \cite{2010PASP..122.1483T} \\			
			2013.5537 & $121.2$ & $0.407$ & -0.2 &  0.004  & \cite{2015ApJ...799....4R} \\
			2014.7629 & $125.0$ & $0.403$ & -0.1 & 0.000 & \cite{2015AJ....150...50T} \\
			2016.961 & $131.9$ & $0.397$ & 0.1 & -0.002 & \cite{2018AJ....156...48T} \\
			\hline
			rms & & & 0.686 & 0.004 & \\
			\hline\hline
		\end{tabular}
		\\
		\medskip
		\end{center}
\end{table}

\subsection{Masses} \label{2.1}

Depending on the calculated orbital elements, the period (P in years) and the semi-major axis (a in arcsec)  in Table~\ref{tabb}, we can employ the latest  Gaia parallax measurement and  the revised \textit{Hipparcos} parallax to calculate the total mass ($\mathcal{M}_{Tot}$) of the system (in solar masses) using Kepler's Third Law as follows:

\begin{equation}
\label{eq31}
\ \mathcal{M}_{Tot} = \frac{a^3}{\varpi^3P^2}\ \mathcal{M}_\odot,\
\end{equation}

\begin{equation}
\frac{\sigma_\mathcal{M} }{\mathcal{M}} =\sqrt{9\Big(\frac{\sigma_\varpi}{\varpi}\Big)^2+9\Big(\frac{\sigma_a}{a}\Big)^2+4\Big(\frac{\sigma_P}{P}\Big)^2}.
\end{equation}

The results are $2.59\pm0.40\mathcal{M}_\odot$ and $2.15\pm0.35\mathcal{M}_\odot$, respectively which will subsequently be compared with the derived masses from the positions of the components on the HR diagram as defined by their physical parameters using Al-Wardat's complex method. 

\begin{table}[ht]
	\begin{center}
		\caption{Orbits, total masses, and quality controls published for the HIP\,109951 system, compared with the orbital solution calculated in this work.}
		\label{tabb}
		\begin{tabular}{c|c|c|c|c|c}
			\noalign{\smallskip}
			\hline\hline
			\noalign{\smallskip}
		&		\multicolumn{4}{c}{System HIP 109951} \\
			\cline{2-6}
			\noalign{\smallskip}
			Parameters	 & \cite{2010AJ....139..205H} & \cite{2014AJ....147...62C} & \cite{2015ApJ...799....4R} &\cite{2018AJ....156...48T} & This work\\
			\hline
			$\rm P$   [yr]       & $134\pm 1.4$ & $80.574\pm 1.145$ & $44\pm 24$ & $50.49\pm 4.32$ & $50.090\pm 2.000$
			\\
			$\rm T_0$   [yr]   & $2128.0 \pm 1.3$ & $1991.763 \pm 1.339$& $1989.0\pm 5.30$ & $1989.88\pm 0.16$& $2040.038 \pm 2.000$
			\\
			$\rm e$     & $0.580 \pm 0.0014$ & $0.450 \pm 0.008$& $0.55\pm 0.59$ &$0.451\pm 0.044$& $0.449 \pm 0.008$
			\\
			$\rm a $  [arcs] & $0.567\pm 0.007 $ & $0.389\pm 0.0042 $  & $0.30\pm 0.16$ & $0.2834\pm 0.020$& $0.282\pm 0.008 $
			\\
			$\rm i $   [deg]    & $ 45.3\pm 2.2$  &  $ 36.0\pm 0.50$ & $36.0\pm 101.0$ & $17.7\pm 13.6$&   $ 15.1\pm 5.5$
			\\
			$\rm \Omega$    [deg] & $248.7 \pm 1.2$  &  $ 71.8\pm 0.80$&  $160.0\pm 62$ & $262.0\pm 3.0$& $267.6\pm 10.0$
			\\
			$\rm \omega$   [deg] & $119.5\pm 3.6$   & $273.9\pm 1.2$ &$126.0\pm 104$ &$45.2\pm 9.6$& $39.2\pm 15.0$
			\\

			rms ($\theta$) [deg] & 0.780 & 1.034 & -& 0.711 & 0.686\\
			rms ($\rho$) [arcsec] & 0.009 & 0.006 &-&  0.005 & 0.004\\
			
			\hline			\hline
			\noalign{\smallskip}
		\end{tabular}
		\\
		\medskip
	\end{center}
\end{table}

\subsection{ Atmospheric modeling}\label{22a}
We begin by estimating the physical properties which involve two parameters: the magnitude difference between the components and the entire visual magnitude, ($ m_v$). The first parameter is taken from speckle interferometric observations while the second parameter is taken from \cite{1997yCat.1239....0E} photometric data.

First, we analyze the system as a binary system $(A-B)$ following the Al-Wardat method using $\Delta m_{A,B} \mathcal = 1.88$ as the average value for the complete list of $\Delta m$ measurements corresponding to the  $545-551$nm $V-band$ filters (see Table~\ref{tab33}). Because the Ba,Bb system is considered as an SB1, it is reasonable to admit that $\Delta m_{Ba,Bb} \geq 1.5$. We will take the real value of 1.5 as magnitude difference between the components of the Ba-Bb system. 
Combining each of the magnitude differences with the entire visual magnitude ($ m_v$) in the Johnson V-band filter (Table~\ref{tab2}) and then with the parallax~ allowed us to calculate the apparent and absolute magnitudes of the individual components ($m^{*}_{v}$ and $M^{*}_{V}$) using the following simple relationships:
\begin{eqnarray}\centering
\label{eq30}
\ m_v^A=m_v+2.5\log(1+10^{-0.4\triangle m}),
\end{eqnarray}
\begin{eqnarray}
\centering
\label{eq32}
\ m_v^B=m_v^A+{\triangle m},
\end{eqnarray}
and
\begin{eqnarray}
\centering
\label{eq3}
\ M_V-m_v &=5-5\log(d)-A_v
\end{eqnarray}
which yield: $m_v^{A}=8\fm90\pm0.05$ and $m_v^{B}=10\fm78$ for the primary and secondary components of the system. Here, we adopt the entire visual magnitude of the Ba-Bb system is the visual apparent magnitude of the component B, so $m_{vBaBb}$ is equal to $10\fm78$. The apparent magnitudes of the Ba-Bb are as follows: $m_v^{Ba}=11\fm02$ and $m_v^{Bb}=12\fm52$ for the primary and secondary of sub-binary components, respectively and  $ M_V^{A}=4\fm61\pm0.09$ and  $M_V^{Ba}=6\fm73\pm0.07$, $M_V^{Bb}=8\fm23\pm0.07$ for the primary and sub-binary components of the triple system, respectively. In the case, we adopted the recently published interstellar absorption for HIP 109951 as $A_{v}=0.19$ \cite{1998ApJ...500..525S} and the distance of the system was estimated using $ d=1/\varpi$ where $\varpi$ is in arcsec, while the error of distance was estimated using $\sigma_d=\sigma_\varpi/\varpi^2$.

As for the errors in Eqs.~\ref{eq30}, ~\ref{eq32}, and \ref{eq3}, they were calculated using the following equations, respectively:
\begin{equation}
\label{eq33}
\sigma_{m^A_{v}} =\pm \sqrt{\sigma_{m_{v}}^2+\Big(\frac{10^{-0.4\triangle m}}{1+10^{-0.4\triangle m}}\Big)^2\sigma_{\triangle m}^2},
\end{equation}

\begin{equation}
\label{eq331}
\sigma_{m^B_{v}} =\pm \sqrt{\sigma_{m^A_{v}}^2+\sigma_{\triangle m}^2},\ and
\end{equation}

\begin{equation}
\label{eq341}
\sigma_{M^{*}_{V}} =\pm \sqrt{\sigma_{m^{*}_{v}}^2+\Big(\frac{5 \log e}{\varpi_{Gaia}}\Big)^2\sigma_{\varpi_{Gaia}}^2}.
\end{equation}

Since the value of the error of visual magnitude, ($\sigma_{m_{v}}$), was extremely small and not given in the photometric data in Table~\ref{tab2}, it can be ignored in Equation~\ref{eq33}.

Starting from the estimated absolute magnitudes of the individual components of the system, and using the relations of absolute magnitude-effective temperature [$M_{V}-T_{\rm eff.}$], mass-luminosity [$\mathcal{M}-L$], spectral type-absolute magnitude [$S_{P}- M_{V}$] \citep{1992adps.book.....L, 2005oasp.book.....G}, and theoretical equations of the main sequence stars:
\begin{eqnarray}
\label{eq8}
\log(R/R_\odot)= 0.5 \log(L/L_\odot)-2\log(T_{\rm eff}/T_\odot),\\
\label{eq25}
\log g = \log(M/M_\odot)- 2\log(R/R_\odot) + 4.43,
\end{eqnarray}
we obtain the preliminary stellar parameters, as follows: $T^{A}_{\rm eff.}=5840K$, log g$^{A}$ = 4.38, $R^{A}=1.16 R_\odot$ for the primary component of the system and $T^{Ba}_{\rm eff.}=5015K$, log g$^{Ba}$ = 4.52, $R^{Ba}= 0.71 R_\odot$ for the primary component of the sub-binary system and $T^{Bb}_{\rm eff.}=4340K$, log g$^{Bb}$ = 4.64, $R^{Bb}= 0.59 R_\odot$ for the secondary component of the sub-binary system . Here, the effective temperature of the Sun was taken as $5777\,{K}$ and its bolometric absolute magnitude as $4\fm75$.

For the sake of analyzing and estimating the physical and geometrical parameters of HIP 109951, we used Al-Wardat's complex method which employs Kurucz Atlas9 line-blanketed grid models~\citep{1994KurCD..19.....K}  to construct individual SEDs and a special subroutine to calculate the entire synthetic SED in order to compare it with the observational one in an iterated manner to obtain the best fit between them.

The entire synthetic SED of the system which is related to the energy flux  of the components located at a distance d (pc) from the Earth according to the following equation is:

\begin{equation}
\label{eq77}
F^{A-B}_{\lambda} \cdot d^2 = H_{\lambda}^{\rm A} \cdot R_{\rm A}^2 +
H_{\lambda}^{\rm B} \cdot R_{\rm B}^2\,
\end{equation}
where
\begin{equation}
\label{eq99}
H_{\lambda}^{\rm B} \cdot R_{\rm B}^2= H_{\lambda}^{\rm Ba} \cdot R_{\rm Ba}^2 +
H_{\lambda}^{\rm Bb} \cdot R_{\rm Bb}^2\,
\end{equation}

 As a result, Equation~\ref{eq77} can be written as,

\begin{equation}
\label{eqFlux}
F_{\lambda}  = (1/d)^2(H_{\lambda}^{\rm A} \cdot R_{\rm A}^2 +
H_{\lambda}^{\rm Ba} \cdot R_{\rm Ba}^2+H_{\lambda}^{\rm Bb} \cdot R_{\rm Bb}^2)
\end{equation}

\noindent where $R_{A}$, $R_{Ba}$, and $R_{Bb}$ are the radii of the components in solar units, $H_\lambda ^{A}$, $H_\lambda ^{Ba}$, and $H_\lambda ^{Bb}$ are the fluxes at the surface of each component star, and $F_\lambda$ is the flux for the entire SED of the system at the surface of the Earth in units of ergs cm$^{-2}$s$^{-1}$ \AA$^{-1}$.

\begin{figure}[ht]
	\includegraphics[angle=0,width=14cm]{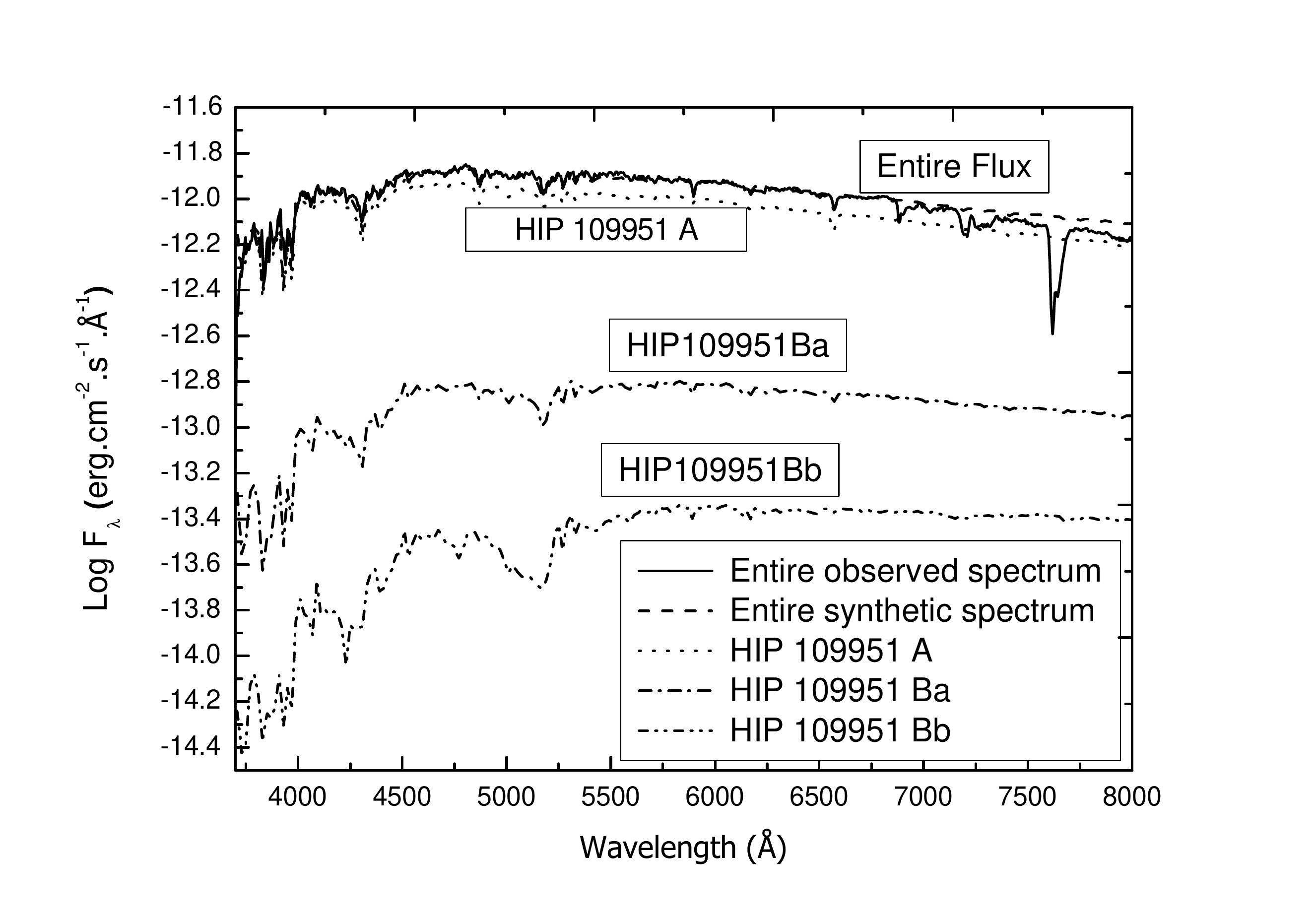}
	\caption{The observed SED of the HIP\,109951 close triple system compared with the synthetic SED using the Kurucz blanketed models ~\citep{1994KurCD..19.....K}. The observed SED is shown with a black solid line.
	}\label{aa2}
\end{figure}

Several tests have been performed on the calculated physical and dynamical parameters. First of all, we compared the entire synthetic SED of the system with the observational spectrum, then we compared the entire synthetic color indices and magnitudes of the system, particularly $(B-V)_{J}$,$\ V_{J}$, and the magnitude difference, $\Delta m$, with the entire observed photometry of the system. Following all of the steps of the fitting and the iteration method of different sets of parameters as well as  the best fit between the color indices of the observed and theoretical synthetic spectra, we obtained the best agreement between the synthetic and observed spectra based on the Gaia parallax \cite{2018yCat.1345....0G} utilizing the following parameters (Figure~\ref{aa2}): $ T_{\rm eff.}^{A}=5836\pm80\,{\rm K}, T_{\rm eff.}^{Ba}=5115\pm80\,{\rm K}, T_{\rm eff.}^{Bb}=4500\pm80\,{\rm K}$, 
$ \log	g_{A}=4.45\pm0.05, \log	g_{Ba}=4.60\pm0.06, \log	g_{Bb}=4.65\pm0.06, $ and 
$R_{A}=1.159\pm0.039\,R_\odot, R_{Ba}=0.634\pm0.05\,R_\odot,$ and $R_{Bb}=0.521\pm0.06\,R_\odot$. On the other hand, we obtained the best agreement between the synthetic and the observed spectra based on the revised \textit{Hipparcos} parallax \cite{2007AA...474..653V} utilizing the same stellar parameters excluding the radii of three components which were as follows (Figure~\ref{aa2}): $R_{A}=1.090\pm0.039\,R_\odot, R_{Ba}=0.596\pm0.05\,R_\odot,$ and $R_{Bb}=0.490\pm0.06\,R_\odot$.

There are two solutions for the close visual triple system, one using the Gaia parallax and the other using the revised \textit{Hipparcos} parallax. But, as a result of the mixed power spread function (PSF) of two components, the Gaia
parallaxes of bright stars (especially close binaries) may be significantly distorted by systematic
errors and the high accuracy
of the Gaia parallax is not final proof of the adequacy of this value of parallax for this star \cite{2016PASP..128l4204L}. That is why, the  solution using the revised \textit{Hipparcos} parallax  \cite{2007AA...474..653V}  is   better   for the HIP 109951 triple system for representing the best physical and dynamical parameters, taking into account the errors of the observed spectrum and magnitudes.

Using the aforementioned parameters of the triple system, we obtain the stellar luminosities of the components: $L_{A}=1.24\pm0.091 L_\odot$, $L_{Ba}=0.22\pm0.09 L_\odot$, and $L_{Bb}=0.09\pm0.05 L_\odot$. Hence, the individual bolometric magnitudes are: $M^{A}_{bol}=4.52\pm0.08$ mag, $M^{Ba}_{bol}=6.39\pm0.08$ mag, and $M^{Bb}_{bol}=7.36\pm0.05$ mag.

Based on the calculated physical and dynamical parameters, specifically the effective temperatures ($T_{\rm eff.}$) and the stellar luminosities ($L$), we estimated the masses using the theoretical evolutionary tracks method with the locations of the components on the H-R diagram (Figure.~\ref{fig3}) as follows:
$\mathcal{M}^{A}=1.05\pm0.16 \mathcal{M}_\odot$, $\mathcal{M}^{Ba}=0.83\pm0.09 \mathcal{M}_\odot$, and $\mathcal{M}^{Bb}=0.67\pm0.04 \mathcal{M}_\odot$. These results allowed us to propose the new spectral types of the triple system as follows: Sp$^{A}\approx$ G1.5, Sp$^{Ba}\approx$ K1.5, and Sp$^{Bb}\approx$ K7.

The total mass of the system was estimated using Al-Wardat's method to be $2.55\mathcal{M}_\odot$,  coinciding to a large extent with those given by Docobo's analytic method as $2.59 \mathcal{M}_\odot$ using the Gaia parallax . 

To estimate the metallicity and ages for HIP 109951, one should use the synthetic isochrones tracks as a functions of metallicity and age given by ~\cite{2000A&AS..141..371G} when $T_{ \text{eff.}}$ and the luminosity of the binary system are known. The initial chemical compositions of all phases of [$Z = 0.0004, Y = 0.23$], [$Z = 0.001, Y = 0.23$], [$Z = 0.004, Y =0.24$], [$Z = 0.008, Y = 0.25$], [$Z = 0.019, Y = 0.273$] (solar composition), and [$Z = 0.03, Y = 0.30$] from the zero-age main sequence star to the first phase of carbon burning. One can see that the metallicity of the system is subject to the positions of the system's components on the synthetic isochrones tracks. As a result, the metallicity for HIP 109951 is $Z=0.008 $ as shown in Fig.~\ref{fig5} and helium of the system is $Y=0.25$, while the system age on the synthetic isochrones tracks as a function of age is between $8.9$ and  $13$\,Gyr.

\subsubsection{Synthetic photometry}\label{1}

Synthetic photometry is used to estimate the stellar parameters more accurately with the color
indices. It is a quantitative analysis for the synthetic SED of a binary system which is used to modify stellar parameters such that the predicted
magnitudes fit the observed ones.
It is predominantly based on the following equation \citep{2006AJ....131.1184M,2007ASPC..364..227M}:

\begin{equation}
\label{55}
m_p[F_{\lambda,s}(\lambda)] = -2.5 \log \frac{\int P_{p}(\lambda)F_{\lambda,s}(\lambda)\lambda{\rm d}\lambda}{\int P_{p}(\lambda)F_{\lambda,r}(\lambda)\lambda{\rm d}\lambda}+ {\rm ZP}_p.
\end{equation}
This equation is crucial for calculating the total and the individual synthetic magnitudes and color indices of the stars. Where $m_p$ is the synthetic magnitude of the $p$ passband, $P_p(\lambda)$ is the dimensionless sensitivity function of the $p$ passband, $F_{\lambda,s}(\lambda)$ is the synthetic SED of the object, and $F_{\lambda,r}(\lambda)$ is the SED of the reference star. Here, zero points (ZP$_p$) were adopted from \cite{2007ASPC..364..227M}.

The result of applying Equation~\ref{55} to the estimated synthetic SED should be consistent with the observational values in Table~\ref{tab2}, otherwise, a new set of parameters should be applied.

Using Equation~\ref{55}, the calculated magnitudes and color indices within three different photometrical systems: Johnson-Cousins: $U$, $B$, $ V$, $R$, $U-B$, $B-V$, $V-R$; Str\"{o}mgren: $u$, $v$, $b$, $y$, $u-v$, $v-b$, $b-y$; and Tycho: $B_{T}$, $ V_{T}$, $B_{T}-V_{T}$ of component $A$ and component $B$ of the entire system $A-B$ are shown in Table~\ref{tab34}.
\begin{table}[ht]
	\begin{center}
		\caption{ Magnitudes and color indices of the composed synthetic spectrum and individual component $A$ and entire of the sub system $B$ of HIP\,109951.}
		\label{tab34}
		\begin{tabular}{l|c|c|c|c}
			\noalign{\smallskip}
			\hline\hline
			\noalign{\smallskip}
			Sys. & Filter & Entire Synth.& HIP\,109951  & HIP\,109951 \\
			&     & $\sigma=\pm0.03$&   A   & B  \\
			\hline
			\noalign{\smallskip}
			Joh-          & $U$ & 9.67 & 9.76 & 12.43   \\
			Cou.          & $B$ & 9.43   &  9.57 & 11.74  \\
			& $V$ & 8.72 &  8.90 & 10.78 \\
			& $R$ & 8.33  &  8.54 & 10.24  \\
			&$U-B$& 0.24  & 0.19 & 0.69  \\
			&$B-V$& 0.71  &  0.67 & 0.96   \\
			&$V-R$& 0.39  &  0.36 & 0.54   \\
			\hline
			\noalign{\smallskip}
			Str\"{o}m.    & $u$ & 10.82 & 10.91 & 13.63  \\
			& $v$ & 9.81 & 9.93  & 12.29  \\
			& $b$ & 9.11 & 9.27 & 11.28  \\
			&  $y$& 8.69 & 8.87 & 10.72  \\
			&$u-v$& 1.01 & 0.98 & 1.34   \\
			&$v-b$& 0.70 & 0.66 & 1.01  \\
			&$b-y$& 0.43 & 0.41 & 0.55  \\
			\hline
			\noalign{\smallskip}
			Tycho       &$B_T$  & 9.61   & 9.74 & 12.00   \\
			&$V_T$  & 8.80   & 8.97 & 10.89  \\
			&$B_T-V_T$& 0.81 & 0.77 & 1.11 \\
			\hline\hline
			\noalign{\smallskip}
		\end{tabular}
	\end{center}
\end{table}

Similarly, the calculated magnitudes and color indices within three different photometrical systems: Johnson-Cousins: $U$, $B$, $ V$, $R$, $U-B$, $B-V$, $V-R$; Str\"{o}mgren: $u$, $v$, $b$, $y$, $u-v$, $v-b$, $b-y$; and Tycho: $B_{T}$, $ V_{T}$, $B_{T}-V_{T}$ of the $Ba$ and $Bb$ components of the sub-system $Ba-Bb$ with entire system are shown in Table~\ref{tab66}.

\begin{table}[ht]
	\begin{center}
		\caption{ Magnitudes and color indices of the composed synthetic spectrum and individual component $Ba-Bb$ and entire of the sub system $B$ of HIP\,109951.}
		\label{tab66}
		\begin{tabular}{l|c|c|c|c}
			\noalign{\smallskip}
			\hline\hline
			\noalign{\smallskip}
			Sys. & Filter & Entire Synth. B& HIP\,109951  & HIP\,109951 \\
			&     & $\sigma=\pm0.03$&   Ba   & Bb  \\
			\hline
			\noalign{\smallskip}
			Joh-          & $U$ & 12.41 & 12.54  & 14.78   \\
			Cou.          & $B$ & 11.74   &  11.93 & 13.71  \\
			& $V$ & 10.78 &  11.02 & 12.52 \\
			& $R$ & 10.23  &  10.52 &  11.80  \\
			&$U-B$& 0.67  &  0.60 & 1.07  \\
			&$B-V$& 0.96  &  0.91 & 1.19   \\
			&$V-R$& 0.55 &  0.50 &  0.72   \\
			\hline
			\noalign{\smallskip}
			Str\"{o}m.    & $u$ & 13.60 & 13.71 & 16.13  \\
			& $v$ & 12.29 & 12.45 &  14.42  \\
			& $b$ & 11.28 & 11.50 & 13.15  \\
			&  $y$& 10.72 & 10.97 & 12.45  \\
			&$u-v$& 1.32 & 1.26 & 1.71   \\
			&$v-b$& 1.00 & 0.95 & 1.27  \\
			&$b-y$& 0.56 & 0.53 & 0.70  \\
			\hline
			\noalign{\smallskip}
			Tycho       &$B_T$  & 12.00   & 12.19 & 14.02   \\
			&$V_T$  & 10.89   & 11.12 &  12.67  \\
			&$B_T-V_T$& 1.11 &  1.06 & 1.35 \\
			\hline\hline
			\noalign{\smallskip}
		\end{tabular}
	\end{center}
\end{table}

\section{RESULTS AND DISCUSSION}
Table~\ref{po2} also shows the residuals, $\Delta \theta$ and $\Delta \rho$, and the rms of the system. Our orbital parameters of the visual system are close to that of \cite{2018AJ....156...48T} but the rms in our orbit improved in theta and in rho more than previous rms.  Moreover, the residuals of the radial velocities obtained with these elements and the parallax of Gaia are of the same order as those that \cite{2018AJ....156...48T} determined in the same manner, that is, using  Tokovinin's visual orbit elements.

Table~\ref{tabb} summarizes the results of the accurate orbital solution of the HIP 109951 triple system  which is plotted in Figure~\ref{2} using Docobo's analytic method. Tokovinin has evaluated the orbital parameters of the main $A-B$ binary system and sub-binary $Ba-Bb$ system. Table~\ref{tabb} shows the orbital parameters of the main binary system. The sub-binary system has an orbital period approximately of  111 days \cite{2018AJ....156...48T}.

Employing magnitude difference measurements of speckle interferometry, the entire observational SED of spectrophotometry along with atmospheric modeling using Al-Wardat's complex method resulted in a precise determination of the complete set of the  physical and geometrical parameters of the main system. These parameters  are presented in Table~\ref{tablef1}.

\begin{table}[ht]
	\begin{center}
		\caption{The physical parameters of the individual components of the HIP\,109951 triple system.}
		\label{tablef1}
		\begin{tabular}{c|c|c|c|c}
			\hline\hline
			&      	& \multicolumn{3}{c}{HIP 109951}  \\
			\cline{3-5}
			Parameters & Units	& HIP\,109951 A &  HIP\,109951 Ba &  HIP\,109951 Bb \\
			\hline
			\noalign{\smallskip}
			$\rm T_{\rm eff}$ {$\pm$ $\sigma_{\rm T_{\rm eff}}$}& [~K~] & $5836\pm80$ & $5115\pm80$   & $4500\pm80$ \\
			R {$\pm$ $\sigma_{\rm R}$}  & [R$_{\odot}$] & $1.09\pm0.039$ & $0.596\pm0.05$ & $0.49\pm0.06$   \\
			$\log\rm g$ {$\pm$ $\sigma_{\rm log g}$} & [cgs] & $4.45\pm0.06$ & $4.60\pm0.06$  & $4.65\pm0.06$  \\
			$\rm L $ {$\pm$ $\sigma_{\rm L}$} & [$\rm L_\odot$] & $1.24\pm0.10 $  & $0.22\pm0.09$ & $0.09\pm0.05$   \\
			$\rm M_{bol}$ {$\pm$ $\sigma_{\rm M_{bol}}$} & [mag] &  $4.52\pm0.08$ & $6.39\pm0.08$ & $7.36\pm0.09$ \\
			$\mathcal{M}^{a}$ {$\pm$ $\sigma_{\mathcal{M}}$} & [$\mathcal{M}_{\odot}$]& $1.05\pm0.16$ & $ 0.83\pm0.16$ & $ 0.67\pm0.16$  \\
			Sp. Type$^{b}$ & &  G1.5 & K1.5 & K7   \\				
			\hline\hline
			\noalign{\smallskip}
		\end{tabular}
		\\
		\medskip
		{ $^{a}$ See the evolutionary tracks of masses ( 0.7, 0.8,...., 1.1 $M_\odot$) of~\cite{2000yCat..41410371G} (Figure~\ref{fig3}).}
		{ $^{b}$ Spectral types from the tables of \cite{1992adps.book.....L,2005oasp.book.....G} and the $\rm M_{V}-\rm S_{\rm P}$ relation.}	
	\end{center}
\end{table}

Figure~\ref{aa2} shows the best fit between the observed and the synthetic spectrum along with three components (A, Ba, and Bb) of the system using a grid of Kurucz solar metallicity models.



Table~\ref{tab42} shows the final results of the calculated magnitudes and color  indices within three different photometrical systems: Johnson-Cousins: $U$, $B$, $ V$, $R$, $U-B$, $B-V$, $V-R$; Str\"{o}mgren: $u$, $v$, $b$, $y$, $u-v$, $v-b$, $b-y$; and Tycho: $B_{T}$, $ V_{T}$, $B_{T}-V_{T}$ of the entire system and individual components of the triple system. The apparent magnitudes of the individual components ($m^{A}_{v}$, $m^{Ba}_{v}$ and $m^{Bb}_{v}$) of the synthetic photometry are found to be similar to those from the observed photometry.
\begin{table}[ht]
	\begin{center}
		\caption{ Magnitudes and color indices of the composed synthetic spectrum and individual components of HIP\,109951.}
		\label{tab42}
		\begin{tabular}{l|c|c|c|c|c}
			\noalign{\smallskip}
			\hline\hline
			\noalign{\smallskip}
			Sys. & Filter & Entire Synth.& HIP\,109951  & HIP\,109951 & HIP\,109951\\
			&     & $\sigma=\pm0.03$&   A   & Ba  &  Bb  \\
			\hline
			\noalign{\smallskip}
			Joh-          & $U$ & 9.67 & 9.76 & 12.54  & 14.78  \\
			Cou.          & $B$ & 9.43   &  9.57 & 11.93 & 13.71  \\
			& $V$ & 8.72 &  8.90 & 11.02 & 12.52  \\
			& $R$ & 8.33  &  8.54 & 10.52 &  11.80  \\
			&$U-B$& 0.24  & 0.19 & 0.60 & 1.07  \\
			&$B-V$& 0.71  &  0.67 & 0.91 & 1.19  \\
			&$V-R$& 0.39  &  0.36 & 0.50 &  0.72  \\
			\hline
			\noalign{\smallskip}
			Str\"{o}m.    & $u$ & 10.82 & 10.91 & 13.71 & 16.13 \\
			& $v$ & 9.81 & 9.93  & 12.45 &  14.42  \\
			& $b$ & 9.11 & 9.27 & 11.50 & 13.15 \\
			&  $y$& 8.69 & 8.87 & 10.97 & 12.45 \\
			&$u-v$& 1.01 & 0.98 & 1.26 & 1.71  \\
			&$v-b$& 0.70 & 0.66 & 0.95 & 1.27 \\
			&$b-y$& 0.43 & 0.41 & 0.53 & 0.70 \\
			\hline
			\noalign{\smallskip}
			Tycho       &$B_T$  & 9.61   & 9.74 & 12.19 & 14.02   \\
			&$V_T$  & 8.80   & 8.97 & 11.12 &  12.67 \\
			&$B_T-V_T$& 0.81 & 0.77 & 1.06 & 1.35\\
			\hline\hline
			\noalign{\smallskip}
		\end{tabular}
	\end{center}
\end{table}

Moreover, the magnitude difference of the sub-binary obtained from the synthetic photometry in Table~\ref{tab42} is similar to those from the observed photometry.

Table~\ref{synth3} summarizes significant results. It shows a comparison between the  synthetic magnitudes and color indices and the observational ones. Of course, the upshots and comparison led us to a lucid and powerful indication of the reliability of the estimated physical and dynamical parameters of the individual components  of the triple system given in Table~\ref{tablef1}.

\begin{figure}[ht]
	\includegraphics[angle=0,width=13cm]{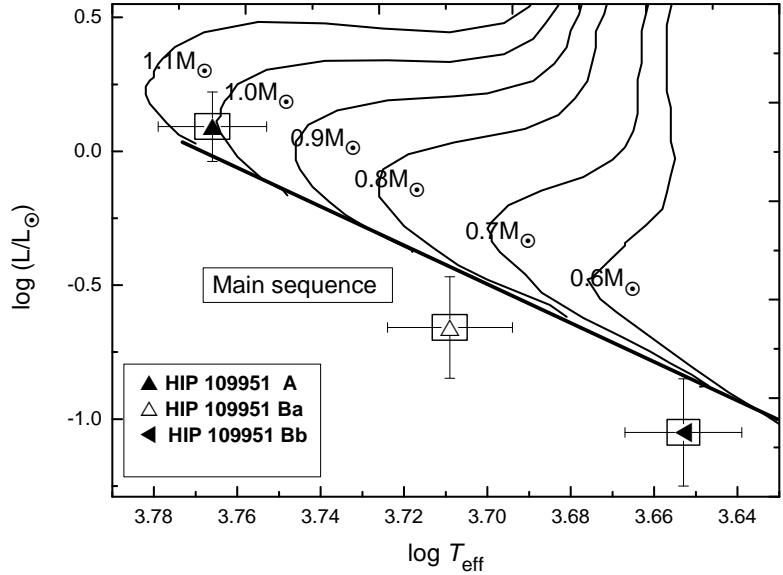}
	\caption{The evolutionary tracks of the components of HIP\,109951 on the H-R diagram of masses ( 0.6, 0.8,...., 1.1 $\mathcal{M}_\odot$), taken from~\cite{2000yCat..41410371G}. }\label{fig3}
\end{figure}

Figure~\ref{fig3} shows the ideal positions of the individual components of the HIP 109951 triple system  following theoretical evolutionary tracks of~\cite{2000yCat..41410371G} on the H-R diagram. This indicates that component A belongs to the main sequence stars, while the components of the sub-binary system are a bit below the main sequence.

We have computed the total mass of the HIP 109951 system using two independent methods: Al-Wardat's complex method and  Docobo's analytical method. The former gave a mass sum of  $2.55\pm0.38\mathcal{M}_\odot$  distributed in the following manner $\mathcal{M}_{A}$ = 1.05 $\mathcal{M}_{\odot}$, $\mathcal{M}_{Ba}$ = 0.83 $\mathcal{M}_{\odot}$, and $\mathcal{M}_{Bb}$ = 0.67 $\mathcal{M}_{\odot}$, while the latter gave a mass sum of $2.15\pm0.35 \mathcal{M}_\odot$ using the revised \textit{Hipparcos} parallax and a mass sum of $2.59\pm0.40 \mathcal{M}_\odot$ using the Gaia parallax. The good agreement in mass sum between the two methods leads to a reliable and accurate analysis of the used methods.

\begin{figure}
	\includegraphics[angle=0,width=13cm]{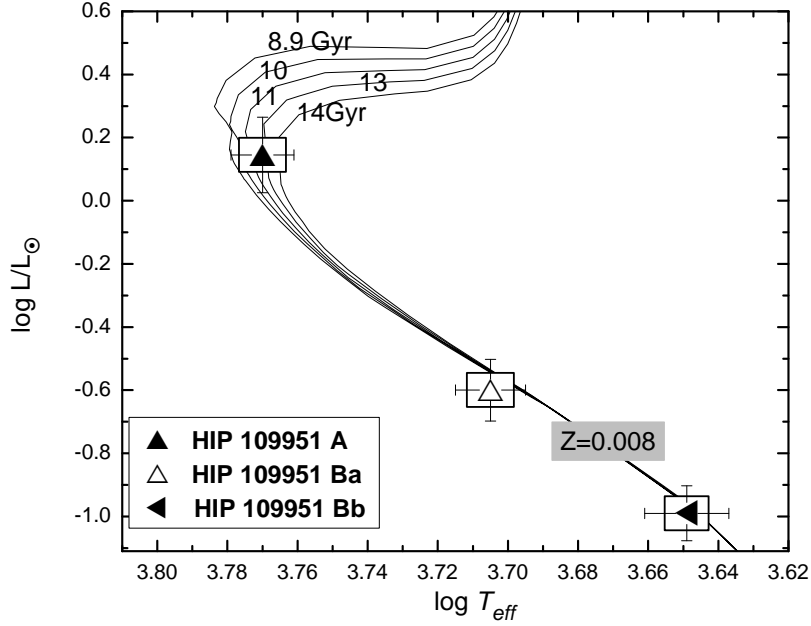}
	\caption{ The isochrones for the components of HIP\,109951 on the H-R diagram for low- and intermediate-mass ranges: from $0.15$ to $7\,\mathcal{M}_\odot$ (and for the composite [Z=0.008, Y=0.25] stars of different metallicities). The isochrones were taken from~\cite{2000AAS..141..371G}. }\label{fig5}
	
\end{figure}

Analysis of HIP\,109951 showed that the system is a slightly metal-deficient star [Z = 0.008, Y = 0.25] (Figure~\ref{fig5}) with an age between $8.9$ and  $13$\,Gyr.

\begin{table}[ht]
	\begin{center}
		\caption{Comparison between the observational and synthetic entire
			magnitudes and color indices of the system HIP\,109951.} \label{synth3}
		\begin{tabular}{c|c|c}
			\noalign{\smallskip}
			\hline
			\hline \noalign{\smallskip}
			&\multicolumn{2}{c}{HIP 109951} \\
			\cline{2-3}
			\noalign{\smallskip}
			Filter	& Observed $^a$ & Synthetic$^b$(This work) \\
			& ($\rm mag$) & ($\rm mag$) \\
			\hline
			\noalign{\smallskip}
			$V_{J}$ & $8.72$ & $8.72\pm0.03$ \\
			$B_J$& $9.43\pm0.03$ & $9.43\pm0.03$  \\
			$B_T$  & $9.58\pm0.03$   &$9.61\pm0.03$ \\
			$V_T$  & $8.82\pm0.02$   &$8.80\pm0.03$ \\
			$(B-V)_{J}$&$ 0.71\pm0.002$ &$ 0.71\pm0.03$ \\
			$(u-v)_{S}$&$ 0.96\pm0.003$ &$ 1.01\pm0.03$ \\
			$(v-b)_{S}$&$ 0.68\pm0.003$ &$ 0.70\pm0.03$ \\
			$(b-y)_{S}$&$ 0.44\pm0.003$ &$ 0.43\pm0.03$ \\
			$\Delta m_{A-B}$ &$1.88\pm0.003$ &$1.88\pm0.003$\\
			$\Delta m{Ba-Bb}$ & $1.50$  &  $1.50\pm 0.003$\\
			\hline\hline
		\end{tabular}
		\\
		\medskip
		{$^a$ Real observations (see Table~\ref{tab2})}
		{$^b$ Entire synthetic work of the HIP 109951 system (see Table~\ref{tab42})
.\\
		}
		
		\end{center}
\end{table}

\section{Conclusions}
Our calculations yielded the best physical and dynamical parameters to date of the studied triple system with reliablity of the parallaxes in both methods.

The total dynamical masses were $\sum \mathcal{M}= $ $2.59\pm0.40\mathcal{M}_\odot$ and $\sum \mathcal{M}= $ $2.15\pm0.35\mathcal{M}_\odot$ using Docobo's method based on the Gaia parallax and  the revised \textit{Hipparcos} parallax, respectively. The total mass of the three components of the system were  $\sum \mathcal{M}= $ $2.55\pm0.38\mathcal{M}_\odot$ using Al-Wardat's method.

In addition, the best match between the observed photometry and synthetic photometry of the magnitudes and color indices in different photometric systems (Johnson: $B$, $ V$, $B-V$; Str\"{o}mgren: $u-v$, $v-b$, $b-y$; and Tycho: $B_{T}$, $ V_{T}$) of the entire system  was accurately achieved.


\section*{Acknowledgments}
This research has made use of SAO/NASA, the SIMBAD database, the Fourth Catalog of Interferometric Measurements of Binary Stars, IPAC data systems, Al-Wardat's code for analyzing close visual binary stars, the CHORIZOS code for photometric and spectrophotometric data analysis.

The authors thank Professor Mashhoor Al-Wardat for his valuable comments and suggestions.

This paper was supported by the Spanish ``Ministerio de Econom\'{i}a, Industria y Competitividad'' under the Project AYA-2016-80938-P (AEI/FEDER, UE) and by the ``Xunta de Galicia'' under the Rede IEMath-Galicia, ED341DR 2016/022 grant.

\bibliographystyle{mnras}
\bibliography{references}

\begin{thebibliography}{}
\makeatletter
\relax
\def\mn@urlcharsother{\let\do\@makeother \do\$\do\&\do\#\do\^\do\_\do\%\do\~}
\def\mn@doi{\begingroup\mn@urlcharsother \@ifnextchar [ {\mn@doi@}
  {\mn@doi@[]}}
\def\mn@doi@[#1]#2{\def\@tempa{#1}\ifx\@tempa\@empty \href
  {http://dx.doi.org/#2} {doi:#2}\else \href {http://dx.doi.org/#2} {#1}\fi
  \endgroup}
\def\mn@eprint#1#2{\mn@eprint@#1:#2::\@nil}
\def\mn@eprint@arXiv#1{\href {http://arxiv.org/abs/#1} {{\tt arXiv:#1}}}
\def\mn@eprint@dblp#1{\href {http://dblp.uni-trier.de/rec/bibtex/#1.xml}
  {dblp:#1}}
\def\mn@eprint@#1:#2:#3:#4\@nil{\def\@tempa {#1}\def\@tempb {#2}\def\@tempc
  {#3}\ifx \@tempc \@empty \let \@tempc \@tempb \let \@tempb \@tempa \fi \ifx
  \@tempb \@empty \def\@tempb {arXiv}\fi \@ifundefined
  {mn@eprint@\@tempb}{\@tempb:\@tempc}{\expandafter \expandafter \csname
  mn@eprint@\@tempb\endcsname \expandafter{\@tempc}}}

\bibitem[\protect\citeauthoryear{{Al-Wardat}}{{Al-Wardat}}{2002}]{2002BSAO...53...51A}
{Al-Wardat} M.~A.,  2002, Bull.~Special Astrophys.~Obs., \href
  {http://adsabs.harvard.edu/abs/2002BSAO...53...51A} {53, 51}

\bibitem[\protect\citeauthoryear{{Al-Wardat}}{{Al-Wardat}}{2003}]{2003BSAO...55...18A}
{Al-Wardat} M.~A.,  2003, Bulletin of the Special Astrophysics Observatory,
  \href {http://adsabs.harvard.edu/abs/2003BSAO...55...18A} {55, 18}

\bibitem[\protect\citeauthoryear{{Al-Wardat}}{{Al-Wardat}}{2007}]{2007AN....328...63A}
{Al-Wardat} M.~A.,  2007, \mn@doi [Astronomische Nachrichten]
  {10.1002/asna.200610676}, \href
  {http://adsabs.harvard.edu/abs/2007AN....328...63A} {328, 63}

\bibitem[\protect\citeauthoryear{{Al-Wardat}}{{Al-Wardat}}{2009}]{2009AN....330..385A}
{Al-Wardat} M.~A.,  2009, Astronomische Nachrichten, \href
  {http://adsabs.harvard.edu/abs/2009AN....330..385A} {330, 385}

\bibitem[\protect\citeauthoryear{{Al-Wardat}}{{Al-Wardat}}{2012}]{2012PASA...29..523A}
{Al-Wardat} M.,  2012, \mn@doi [\pasa] {10.1071/AS12004}, \href
  {https://ui.adsabs.harvard.edu/abs/2012PASA...29..523A} {29, 523}

\bibitem[\protect\citeauthoryear{{Al-Wardat}}{{Al-Wardat}}{2014}]{2014AstBu..69..454A}
{Al-Wardat} M.~A.,  2014, \mn@doi [Astrophysical Bulletin]
  {10.1134/S1990341314040075}, \href
  {https://ui.adsabs.harvard.edu/abs/2014AstBu..69..454A} {69, 454}

\bibitem[\protect\citeauthoryear{{Al-Wardat} \& {Widyan}}{{Al-Wardat} \&
  {Widyan}}{2009}]{2009AstBu..64..365A}
{Al-Wardat} M.~A.,  {Widyan} H.,  2009, Astrophysical Bulletin, \href
  {http://adsabs.harvard.edu/abs/2009AstBu..64..365A} {64, 365}

\bibitem[\protect\citeauthoryear{{Al-Wardat}, {Widyan}  \&
  {Al-thyabat}}{{Al-Wardat} et~al.}{2014a}]{2014PASA...31....5A}
{Al-Wardat} M.~A.,  {Widyan} H.~S.,   {Al-thyabat} A.,  2014a, \mn@doi [\pasa]
  {10.1017/pasa.2013.42}, \href
  {http://adsabs.harvard.edu/abs/2014PASA...31....5A} {31, e005}

\bibitem[\protect\citeauthoryear{{Al-Wardat}, {Balega}, {Leushin}, {Yusuf},
  {Taani}, {Al-Waqfi}  \& {Masda}}{{Al-Wardat}
  et~al.}{2014b}]{2014AstBu..69...58A}
{Al-Wardat} M.~A.,  {Balega} Y.~Y.,  {Leushin} V.~V.,  {Yusuf} N.~A.,  {Taani}
  A.~A.,  {Al-Waqfi} K.~S.,   {Masda} S.,  2014b, Astrophysical Bulletin, \href
  {http://adsabs.harvard.edu/abs/2014AstBu..69...58A} {69, 58}

\bibitem[\protect\citeauthoryear{{Al-Wardat}, {Balega}, {Leushin}, {Zuchkov},
  {Abujbha}, {Al-Waqfi}  \& {Masda}}{{Al-Wardat}
  et~al.}{2014c}]{2014AstBu..69..198A}
{Al-Wardat} M.~A.,  {Balega} Y.~Y.,  {Leushin} V.~V.,  {Zuchkov} R.~Y.,
  {Abujbha} R.~M.,  {Al-Waqfi} K.~S.,   {Masda} S.,  2014c, \mn@doi
  [Astrophysical Bulletin] {10.1134/S1990341314020072}, \href
  {http://adsabs.harvard.edu/abs/2014AstBu..69..198A} {69, 198}

\bibitem[\protect\citeauthoryear{{Al-Wardat}, {El-Mahameed}, {Yusuf},
  {Khasawneh}  \& {Masda}}{{Al-Wardat} et~al.}{2016}]{2016RAA....16..166A}
{Al-Wardat} M.~A.,  {El-Mahameed} M.~H.,  {Yusuf} N.~A.,  {Khasawneh} A.~M.,
  {Masda} S.~G.,  2016, \mn@doi [Research in Astronomy and Astrophysics]
  {10.1088/1674-4527/16/11/166}, \href
  {http://adsabs.harvard.edu/abs/2016RAA....16..166A} {16, 166}

\bibitem[\protect\citeauthoryear{{Al-Wardat}, {Docobo}, {Abushattal}  \&
  {Campo}}{{Al-Wardat} et~al.}{2017}]{2017AstBu..72...24A}
{Al-Wardat} M.~A.,  {Docobo} J.~A.,  {Abushattal} A.~A.,   {Campo} P.~P.,
  2017, \mn@doi [Astrophysical Bulletin] {10.1134/S1990341317030038}, \href
  {http://adsabs.harvard.edu/abs/2017AstBu..72...24A} {72, 24}

\bibitem[\protect\citeauthoryear{{Balega}, {Balega}, {Maksimov}, {Pluzhnik},
  {Schertl}, {Shkhagosheva}  \& {Weigelt}}{{Balega}
  et~al.}{2004}]{2004AA...422..627B}
{Balega} I.,  {Balega} Y.~Y.,  {Maksimov} A.~F.,  {Pluzhnik} E.~A.,  {Schertl}
  D.,  {Shkhagosheva} Z.~U.,   {Weigelt} G.,  2004, \aap, \href
  {http://adsabs.harvard.edu/abs/2004A%26A...422..627B} {422, 627}

\bibitem[\protect\citeauthoryear{{Balega}, {Balega}, {Maksimov},
  {Malogolovets}, {Pluzhnik}  \& {Shkhagosheva}}{{Balega}
  et~al.}{2006}]{2006BSAO...59...20B}
{Balega} I.~I.,  {Balega} A.~F.,  {Maksimov} E.~V.,  {Malogolovets} E.~A.,
  {Pluzhnik} E.~A.,   {Shkhagosheva} Z.~U.,  2006, Bull.~Special
  Astrophys.~Obs., \href {http://adsabs.harvard.edu/abs/2006BSAO...59...20B}
  {59, 20}

\bibitem[\protect\citeauthoryear{{Balega}, {Balega}, {Maksimov},
  {Malogolovets}, {Rastegaev}, {Shkhagosheva}  \& {Weigelt}}{{Balega}
  et~al.}{2007}]{2007AstBu..62..339B}
{Balega} I.~I.,  {Balega} Y.~Y.,  {Maksimov} A.~F.,  {Malogolovets} E.~V.,
  {Rastegaev} D.~A.,  {Shkhagosheva} Z.~U.,   {Weigelt} G.,  2007,
  Astrophysical Bulletin, \href
  {http://adsabs.harvard.edu/abs/2007AstBu..62..339B} {62, 339}

\bibitem[\protect\citeauthoryear{{Cvetkovi{\'c}}, {Pavlovi{\'c}}  \&
  {Ninkovi{\'c}}}{{Cvetkovi{\'c}} et~al.}{2014}]{2014AJ....147...62C}
{Cvetkovi{\'c}} Z.,  {Pavlovi{\'c}} R.,   {Ninkovi{\'c}} S.,  2014, \mn@doi
  [\aj] {10.1088/0004-6256/147/3/62}, \href
  {http://adsabs.harvard.edu/abs/2014AJ....147...62C} {147, 62}

\bibitem[\protect\citeauthoryear{{Docobo}}{{Docobo}}{1985}]{1985CeMec..36..143D}
{Docobo} J.~A.,  1985, Celestial Mechanics, \href
  {http://adsabs.harvard.edu/abs/1985CeMec..36..143D} {36, 143}

\bibitem[\protect\citeauthoryear{{Docobo}}{{Docobo}}{2012}]{2012ocpd.conf..119D}
{Docobo} J.~A.,  2012, in {Arenou} F.,  {Hestroffer} D.,  eds, Orbital Couples:
  Pas de Deux in the Solar System and the Milky Way. pp 119--123

\bibitem[\protect\citeauthoryear{{Docobo}, {Tamazian}  \& {Campo}}{{Docobo}
  et~al.}{2018}]{2018MNRAS.476.2792D}
{Docobo} J.~A.,  {Tamazian} V.~S.,   {Campo} P.~P.,  2018, \mn@doi [\mnras]
  {10.1093/mnras/sty317}, \href
  {http://adsabs.harvard.edu/abs/2018MNRAS.476.2792D} {476, 2792}

\bibitem[\protect\citeauthoryear{{ESA}}{{ESA}}{1997}]{1997yCat.1239....0E}
{ESA} 1997, {The Hipparcos and Tycho Catalogues (ESA)}

\bibitem[\protect\citeauthoryear{{Gaia Collaboration}}{{Gaia
  Collaboration}}{2018}]{2018yCat.1345....0G}
{Gaia Collaboration} 2018, VizieR Online Data Catalog, \href
  {http://adsabs.harvard.edu/abs/2018yCat.1345....0G} {1345}

\bibitem[\protect\citeauthoryear{{Ghezzi} \& {Johnson}}{{Ghezzi} \&
  {Johnson}}{2015}]{2015ApJ...812...96G}
{Ghezzi} L.,  {Johnson} J.~A.,  2015, \mn@doi [\apj]
  {10.1088/0004-637X/812/2/96}, \href
  {http://adsabs.harvard.edu/abs/2015ApJ...812...96G} {812, 96}

\bibitem[\protect\citeauthoryear{{Girardi}, {Bressan}, {Bertelli}  \&
  {Chiosi}}{{Girardi} et~al.}{2000a}]{2000AAS..141..371G}
{Girardi} L.,  {Bressan} A.,  {Bertelli} G.,   {Chiosi} C.,  2000a, \mn@doi
  [\aaps] {10.1051/aas:2000126}, \href
  {http://adsabs.harvard.edu/abs/2000A%26AS..141..371G} {141, 371}

\bibitem[\protect\citeauthoryear{{Girardi}, {Bressan}, {Bertelli}  \&
  {Chiosi}}{{Girardi} et~al.}{2000b}]{2000A&AS..141..371G}
{Girardi} L.,  {Bressan} A.,  {Bertelli} G.,   {Chiosi} C.,  2000b, \mn@doi
  [\aaps] {10.1051/aas:2000126}, \href
  {https://ui.adsabs.harvard.edu/abs/2000A%26AS..141..371G} {141, 371}

\bibitem[\protect\citeauthoryear{{Girardi}, {Bressan}, {Bertelli}  \&
  {Chiosi}}{{Girardi} et~al.}{2000c}]{2000yCat..41410371G}
{Girardi} L.,  {Bressan} A.,  {Bertelli} G.,   {Chiosi} C.,  2000c, VizieR
  Online Data Catalog, \href
  {http://adsabs.harvard.edu/abs/2000yCat..41410371G} {414, 10371}

\bibitem[\protect\citeauthoryear{{Gray}}{{Gray}}{2005}]{2005oasp.book.....G}
{Gray} D.~F.,  2005, {The Observation and Analysis of Stellar Photospheres}

\bibitem[\protect\citeauthoryear{{Hartkopf}, {McAlister}  \&
  {Mason}}{{Hartkopf} et~al.}{2001}]{2001AJ....122.3480H}
{Hartkopf} W.~I.,  {McAlister} H.~A.,   {Mason} B.~D.,  2001, \aj, \href
  {http://adsabs.harvard.edu/abs/2001AJ....122.3480H} {122, 3480}

\bibitem[\protect\citeauthoryear{{Hauck} \& {Mermilliod}}{{Hauck} \&
  {Mermilliod}}{1998}]{1998AAS..129..431H}
{Hauck} B.,  {Mermilliod} M.,  1998, \mn@doi [\aaps] {10.1051/aas:1998195},
  \href {http://adsabs.harvard.edu/abs/1998A%26AS..129..431H} {129, 431}

\bibitem[\protect\citeauthoryear{{H{\o}g} et~al.,}{{H{\o}g}
  et~al.}{2000}]{2000AA...355L..27H}
{H{\o}g} E.,  et~al., 2000, \aap, \href
  {http://cdsads.u-strasbg.fr/abs/2000A%26A...355L..27H} {355, L27}

\bibitem[\protect\citeauthoryear{{Horch}, {Robinson}, {Meyer}, {van Altena},
  {Ninkov}  \& {Piterman}}{{Horch} et~al.}{2002}]{2002AJ....123.3442H}
{Horch} E.~P.,  {Robinson} S.~E.,  {Meyer} R.~D.,  {van Altena} W.~F.,
  {Ninkov} Z.,   {Piterman} A.,  2002, \aj, \href
  {http://adsabs.harvard.edu/abs/2002AJ....123.3442H} {123, 3442}

\bibitem[\protect\citeauthoryear{{Horch}, {Meyer}  \& {van Altena}}{{Horch}
  et~al.}{2004}]{2004AJ....127.1727H}
{Horch} E.~P.,  {Meyer} R.~D.,   {van Altena} W.~F.,  2004, \aj, \href
  {http://adsabs.harvard.edu/abs/2004AJ....127.1727H} {127, 1727}

\bibitem[\protect\citeauthoryear{{Horch}, {van Altena}, {Cyr}, {Kinsman-Smith},
  {Srivastava}  \& {Zhou}}{{Horch} et~al.}{2008}]{2008AJ....136..312H}
{Horch} E.~P.,  {van Altena} W.~F.,  {Cyr} Jr. W.~M.,  {Kinsman-Smith} L.,
  {Srivastava} A.,   {Zhou} J.,  2008, \aj, \href
  {http://adsabs.harvard.edu/abs/2008AJ....136..312H} {136, 312}

\bibitem[\protect\citeauthoryear{{Horch}, {Falta}, {Anderson}, {DeSousa},
  {Miniter}, {Ahmed}  \& {van Altena}}{{Horch}
  et~al.}{2010}]{2010AJ....139..205H}
{Horch} E.~P.,  {Falta} D.,  {Anderson} L.~M.,  {DeSousa} M.~D.,  {Miniter}
  C.~M.,  {Ahmed} T.,   {van Altena} W.~F.,  2010, \aj, \href
  {http://adsabs.harvard.edu/abs/2010AJ....139..205H} {139, 205}

\bibitem[\protect\citeauthoryear{{Horch}, {Bahi}, {Gaulin}, {Howell}, {Sherry},
  {Baena Gall{\'e}}  \& {van Altena}}{{Horch}
  et~al.}{2012}]{2012AJ....143...10H}
{Horch} E.~P.,  {Bahi} L.~A.~P.,  {Gaulin} J.~R.,  {Howell} S.~B.,  {Sherry}
  W.~H.,  {Baena Gall{\'e}} R.,   {van Altena} W.~F.,  2012, \aj, \href
  {http://adsabs.harvard.edu/abs/2012AJ....143...10H} {143, 10}

\bibitem[\protect\citeauthoryear{{Kurucz}}{{Kurucz}}{1994}]{1994KurCD..19.....K}
{Kurucz} R.,  1994, Solar abundance model atmospheres for 0,1,2,4,8
  km/s.~Kurucz CD-ROM No.~19.~ Cambridge, Mass.: Smithsonian Astrophysical
  Observatory, 1994., \href {http://adsabs.harvard.edu/abs/1994KurCD..19.....K}
  {19}

\bibitem[\protect\citeauthoryear{{Lang}}{{Lang}}{1992}]{1992adps.book.....L}
{Lang} K.~R.,  1992, {Astrophysical Data I. Planets and Stars.}

\bibitem[\protect\citeauthoryear{{Lund} et~al.,}{{Lund}
  et~al.}{2016}]{2016PASP..128l4204L}
{Lund} M.~N.,  et~al., 2016, \mn@doi [\pasp]
  {10.1088/1538-3873/128/970/124204}, \href
  {https://ui.adsabs.harvard.edu/abs/2016PASP..128l4204L} {128, 124204}

\bibitem[\protect\citeauthoryear{{MacKnight} \& {Horch}}{{MacKnight} \&
  {Horch}}{2004}]{2004AAS...204.0719M}
{MacKnight} M.,  {Horch} E.~P.,  2004, in American Astronomical Society Meeting
  Abstracts \#204. p.~788

\bibitem[\protect\citeauthoryear{{Ma{\'{\i}}z Apell{\'a}niz}}{{Ma{\'{\i}}z
  Apell{\'a}niz}}{2006}]{2006AJ....131.1184M}
{Ma{\'{\i}}z Apell{\'a}niz} J.,  2006, \aj, \href
  {http://adsabs.harvard.edu/abs/2006AJ....131.1184M} {131, 1184}

\bibitem[\protect\citeauthoryear{{Ma{\'{\i}}z Apell{\'a}niz}}{{Ma{\'{\i}}z
  Apell{\'a}niz}}{2007}]{2007ASPC..364..227M}
{Ma{\'{\i}}z Apell{\'a}niz} J.,  2007, in {Sterken} C.,  ed.,  Astronomical
  Society of the Pacific Conference Series Vol. 364, The Future of Photometric,
  Spectrophotometric and Polarimetric Standardization. San Francisco:
  Astronomical Society of the Pacific, pp 227--236

\bibitem[\protect\citeauthoryear{{Malkov}, {Sichevskij}  \&
  {Kovaleva}}{{Malkov} et~al.}{2010}]{2010MNRAS.401..695M}
{Malkov} O.~Y.,  {Sichevskij} S.~G.,   {Kovaleva} D.~A.,  2010, \mn@doi
  [\mnras] {10.1111/j.1365-2966.2009.15696.x}, \href
  {http://adsabs.harvard.edu/abs/2010MNRAS.401..695M} {401, 695}

\bibitem[\protect\citeauthoryear{{Masda}, {Al-Wardat}  \& {Pathan}}{{Masda}
  et~al.}{2018a}]{2018arXiv180203804M}
{Masda} S.~G.,  {Al-Wardat} M.~A.,   {Pathan} J.~M.,  2018a, preprint, \href
  {http://adsabs.harvard.edu/abs/2018arXiv180203804M} {} (\mn@eprint {arXiv}
  {1802.03804})

\bibitem[\protect\citeauthoryear{{Masda}, {Al-Wardat}  \& {Pathan}}{{Masda}
  et~al.}{2018b}]{2018JApA...39...58M}
{Masda} S.~G.,  {Al-Wardat} M.~A.,   {Pathan} J.~M.,  2018b, \mn@doi [Journal
  of Astrophysics and Astronomy] {10.1007/s12036-018-9548-z}, \href
  {http://adsabs.harvard.edu/abs/2018JApA...39...58M} {39, 58}

\bibitem[\protect\citeauthoryear{{Pluzhnik}}{{Pluzhnik}}{2005}]{2005AA...431..587P}
{Pluzhnik} E.~A.,  2005, \aap, \href
  {http://adsabs.harvard.edu/abs/2005A26A...431..587P} {431, 587}

\bibitem[\protect\citeauthoryear{{Riddle} et~al.,}{{Riddle}
  et~al.}{2015}]{2015ApJ...799....4R}
{Riddle} R.~L.,  et~al., 2015, \mn@doi [\apj] {10.1088/0004-637X/799/1/4},
  \href {http://adsabs.harvard.edu/abs/2015ApJ...799....4R} {799, 4}

\bibitem[\protect\citeauthoryear{{Schlafly} \& {Finkbeiner}}{{Schlafly} \&
  {Finkbeiner}}{2011}]{2011ApJ...737..103S}
{Schlafly} E.~F.,  {Finkbeiner} D.~P.,  2011, \apj, \href
  {http://adsabs.harvard.edu/abs/2011ApJ...737..103S} {737, 103}

\bibitem[\protect\citeauthoryear{{Schlegel}, {Finkbeiner}  \&
  {Davis}}{{Schlegel} et~al.}{1998}]{1998ApJ...500..525S}
{Schlegel} D.~J.,  {Finkbeiner} D.~P.,   {Davis} M.,  1998, \mn@doi [\apj]
  {10.1086/305772}, \href {http://adsabs.harvard.edu/abs/1998ApJ...500..525S}
  {500, 525}

\bibitem[\protect\citeauthoryear{{Tokovinin}}{{Tokovinin}}{2018}]{2018AJ....156...48T}
{Tokovinin} A.,  2018, \mn@doi [\aj] {10.3847/1538-3881/aacb78}, \href
  {http://adsabs.harvard.edu/abs/2018AJ....156...48T} {156, 48}

\bibitem[\protect\citeauthoryear{{Tokovinin}, {Cantarutti}, {Tighe},
  {Schurter}, {van der Bliek}, {Martinez}  \& {Mondaca}}{{Tokovinin}
  et~al.}{2010a}]{2010PASP..122.1483T}
{Tokovinin} A.,  {Cantarutti} R.,  {Tighe} R.,  {Schurter} P.,  {van der Bliek}
  N.,  {Martinez} M.,   {Mondaca} E.,  2010a, \mn@doi [\pasp] {10.1086/657903},
  \href {http://adsabs.harvard.edu/abs/2010PASP..122.1483T} {122, 1483}

\bibitem[\protect\citeauthoryear{{Tokovinin}, {Mason}  \&
  {Hartkopf}}{{Tokovinin} et~al.}{2010b}]{2010AJ....139..743T}
{Tokovinin} A.,  {Mason} B.~D.,   {Hartkopf} W.~I.,  2010b, \aj, \href
  {http://adsabs.harvard.edu/abs/2010AJ....139..743T} {139, 743}

\bibitem[\protect\citeauthoryear{{Tokovinin}, {Mason}, {Hartkopf}, {Mendez}  \&
  {Horch}}{{Tokovinin} et~al.}{2015}]{2015AJ....150...50T}
{Tokovinin} A.,  {Mason} B.~D.,  {Hartkopf} W.~I.,  {Mendez} R.~A.,   {Horch}
  E.~P.,  2015, \mn@doi [\aj] {10.1088/0004-6256/150/2/50}, \href
  {http://adsabs.harvard.edu/abs/2015AJ....150...50T} {150, 50}

\bibitem[\protect\citeauthoryear{{van Leeuwen}}{{van
  Leeuwen}}{2007}]{2007AA...474..653V}
{van Leeuwen} F.,  2007, \mn@doi [\aap] {10.1051/0004-6361:20078357}, \href
  {http://adsabs.harvard.edu/abs/2007A%26A...474..653V} {474, 653}

\makeatother
\end{thebibliography}


\end{document}